\documentclass[twocolumn,superscriptaddress,showpacs,pra,aps,a4paper,floatfix,10pt,longbibliography]{revtex4-1}
\usepackage{amssymb, amsbsy, amsmath, latexsym, dsfont, array, layout, graphicx,mathrsfs,color}
\newcommand{\fig}[1]{Fig.~\ref{#1}}

\newcommand{\ket}[1]{\left|{#1}\right\rangle}

\usepackage{bbm}

\newcommand{\eq}[1]{Eq.~(\ref{#1})}
\newcommand{\beq}{\begin{eqnarray}}
\newcommand{\eeq}{\end{eqnarray}}
\newcommand{\ew}[1]{\langle #1 \rangle}

\usepackage[normalem]{ulem}
\usepackage{color}

\begin{document}

\title{Violations of a Leggett-Garg inequality without signalling for a photonic qutrit probed with ambiguous measurements}
\author{Kunkun Wang}
\affiliation{Department of Physics, Southeast University, Nanjing 211189, China}
\author{Clive Emary}
\affiliation{Joint Quantum Centre (JQC) Durham-Newcastle, School of Mathematics, Statistics and Physics, Newcastle University, Newcastle-upon-Tyne, NE1 7RU, United Kingdom}
\author{Mengyan Xu}
\affiliation{Department of Physics, Southeast University, Nanjing 211189, China}
\author{Xiang Zhan}
\affiliation{Department of Physics, Southeast University, Nanjing 211189, China}
\author{Zhihao Bian}
\affiliation{Department of Physics, Southeast University, Nanjing 211189, China}
\author{Lei Xiao}
\affiliation{Department of Physics, Southeast University, Nanjing 211189, China}
\author{Peng Xue}
\email{gnep.eux@gmail.com}
\affiliation{Department of Physics, Southeast University, Nanjing 211189, China}
\affiliation{Beijing Computational Science Research Center, Beijing 100084, China}
\affiliation{State Key Laboratory of Precision Spectroscopy, East China Normal University, Shanghai 200062, China}

\begin{abstract}
  We realise a quantum three-level system with photons distributed among three different spatial and polarization modes.  Ambiguous measurement of the state of the qutrit are realised by blocking one out for the three modes at any one time.  Using these measurements we construct a test of a Leggett-Garg inequality as well as tests of no-signalling-in-time  for the measurements.  We observe violations of the Leggett-Garg inequality that can not be accounted for in terms of signalling.  Moreover, we  tailor the qutrit dynamics such that both ambiguous and unambiguous measurements are simultaneously non-signalling, which is an essential step for the justification of the use of ambiguous measurements in Leggett-Garg tests.
\end{abstract}

\maketitle

\section{Introduction}

Macrorealism, as defined by Leggett and Garg~\cite{LG1985}, posits that a macroscopic system will exist in a well-defined state at all times, and that this state can be measured without disturbing it (the assumption of non-invasive measurability).  From these assumptions follow the Leggett-Garg inequalities (LGIs)~\cite{LG1985,Clive2015,Maroney2014}, which hold under macrorealism but can be violated by quantum mechanics \cite{PLaloy2010,Dressel2011,Knee2012,Zhou2015,Robens2015,Knee2016,WCX2017,Katiyar2017}.
The same assumptions also imply the \textit{no-signalling-in-time} (NSIT) equalities, which demonstrate the absence, on the statistical level, of signalling between measurements~\cite{Kofler2013,Li2012,Schild2015,Clemente2015,*Clemente2016}.
Having NSIT hold completes the formal similarity between the temporal LGI and spatial Bell tests \cite{Halliwell2016}.  Violations of a LGI without NSIT, however, provides a convenient loophole for a macrorealist to explain the experiment in terms of the signalling of invasive measurements.

It has been shown theoretically that when unambiguous, projective measurements are used, violations of LGIs are always accompanied by violations of NSIT \cite{George2013,Clive2017}, and thus the use of projective measurements is generally problematic in this context.
In Ref.~\cite{George2013}, however, George \textit{et al.} realised LGI violations without signalling through use of measurements that were \textit{ambiguous}, i.e. measurements where the individual results do not completely reveal the state of the system \cite{Dressel2012}.  Quantum-mechanically, such measurements are sometimes described as ``semi-weak''. LGI violations with ambiguous measurements were also discussed in Refs.~\cite{Dressel2011,Dressel2014,White2016}.
In Ref.~\cite{Clive2017}, a general framework for LGI tests with ambiguous measurements was discussed.  There it was shown that the derivation of LGIs that use data from ambiguous measurements rely on an assumption that equates the invasive influence of the ambiguous measuring device to that of an unambiguous one acting on the same system.
Whilst it is perhaps hard to see how this assumption might hold in general, it has the clear implication that an LGI test in which ambiguous measurements are observed to be non-signalling is only consistent with its own assumptions if the corresponding set of unambiguous measurements on the same system is also observed to be non-signalling.

In this paper, we report on LG experiments with single photons that implement a three-level quantum system measured with both ambiguous and unambiguous  measurements.   We test LGIs and NSIT equalities with both sets of measurements.
In the case of unambiguous measurements, we confirm that all observed LGI violations can explained in terms of signalling.  In the ambiguous case, however, we show that it is possible arrange the time-evolution of our three-level system such that the ambiguously-measured LGI is violated whilst at the same time NSIT is satisfied for both measurement types.
In this case, we obtain an LGI violation that is consistent both with assumption of non-invasive measureability as well as the assumptions implicit in the usage of ambiguous measurements in this type of test.

This paper proceeds as follows.  In Sec.~\ref{SEC:ambig} we describe what is meant here by ambiguous measurements and in Sec.~\ref{SEC:exp} we describe their experimental realisation for our photonic qutrit.  In Sec.~\ref{SEC:uLGI} we consider the non-violations of the LGI with unambiguous measurements when signalling is taken into account.  Section~\ref{SEC:aLGI} contains our main results where we employ ambiguous measurements to violate a LGI whilst all no-signalling constraints are fulfilled.  We conclude with discussions in Sec.~\ref{SEC:disc}.

\begin{figure*}[tb]
   \includegraphics[width=\textwidth]{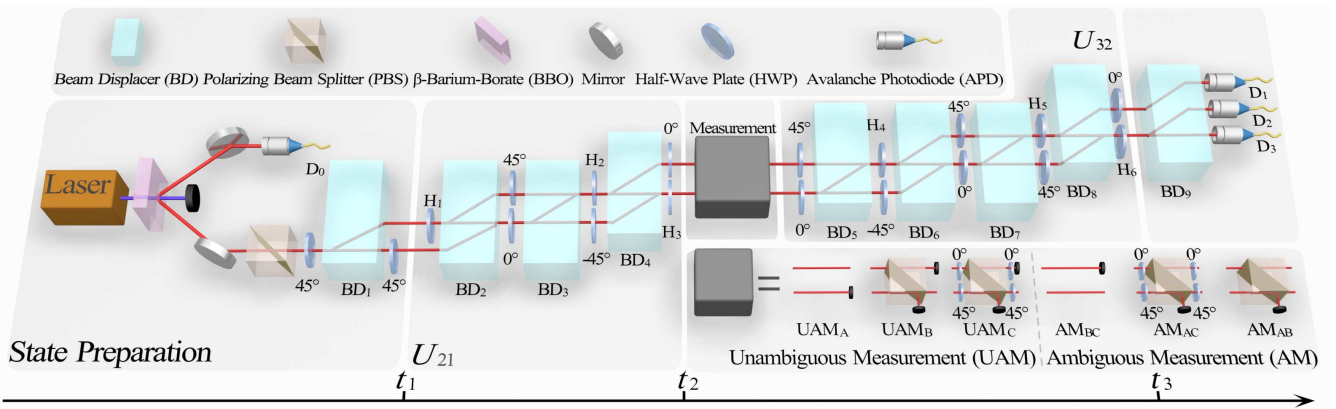}
   \caption{
    Experimental setup for the test of LGI. The heralded single photons are created via type-I spontaneous parametric down-conversion in a $\beta$-barium-borate (BBO) nonlinear crystal and are injected into the optical network (see figure for acronyms). The first polarizing beam splitter (PBS), half-wave plates (HWPs) at $45^\circ$ and BD$_1$ are used to generate the initial qutrit state. The evolution operations $U_{21}$ and $U_{32}$ are realized by HWPs and beam displacers (BDs). The projective measurement at time $t_3$ is realized via the last BD which maps the basis states of the qutrit into three spatial modes. Detecting heralded single photons means in practice registering coincidences between the trigger detector D$_0$ and each of the detectors for measurement D$_1$, D$_2$, and D$_3$.  The unambiguous and ambiguous measurements at time $t_2$ are realized by blocking two or one channels at a time.
   }
\label{FIG:Experimental_setup}
\end{figure*}

\section{Ambiguous measurements \label{SEC:ambig}}

We begin by discussing the meaning of unambiguous and ambiguous measurements following Ref.~\cite{Dressel2012}, which establishes these concepts identically in both quantum and classical contexts.

In our three-level system, unambiguous measurements reveal one of three distinct results $n\in \left\{A,B,C\right\}$, and since these results are repeatable, we associate $n$ with the ``realistic'' system state.  Let us denote the probability that we measure result $n$ as $P(n)$.

On the other hand, ambiguous measurements do not reveal complete information about the state of the system and are non-repeatable.  The particular scheme we will consider here is a set of three individual measurements, each of which serves to exclude one of the three system states.  Thus, the measurement outcomes are $\alpha\in\{B\cup C,A \cup C, A \cup B\}$ and our experiments record probabilities  such as $P(B\cup C)$, etc.

The probabilities obtained with the two different measurements setups are clearly related. Elementary probability theory gives $P(B\cup C) = P(B) + P(C)$, for example, which could easily be verified experimentally.  Given the complete set of three ambiguous probabilities, a macrorealist would have no qualms inferring the probabilities that the system ``really was in'' such as $A$, by calculating
\beq
  P'(A)
  =
  \textstyle{\frac{1}{2}}P(A\cup B)
  +\textstyle{\frac{1}{2}}P(A\cup C)
  -\textstyle{\frac{1}{2}}P(B\cup C)
  \label{EQ:inferPA}
  ,
\eeq
and so on \footnote{This decomposition of a rank-$1$ probability into a sum of rank-$2$ probabilities implicitly relies upon the assumption of the existence of a joint distribution for all three outcomes~\cite{Dressel2011}.}.
Here we maintain the notation $P'$ for a probability inferred from ambiguous measurements, as opposed to one that is measured directly.

Quantum-mechanically, unambiguous and ambiguous measurement are realised respectively as a complete set of projection operators and a more general POVM that implements a ``semi-weak'' measurement \footnote{Note that  notions of ``projective'' and ``semi-weak'' measurements arise from within the formalism of quantum mechanics, and thus they would have  no meaning to a macrorealist and can not be used to specify an LGI test.}.
In the case of a measurement of the systems state at a single time, calculating either with quantum-mechanics or classically, $P'(A)$ and $P(A)$ will clearly agree.   However,  when sequential measurements are made on the same system, the difference  in the quantum case between directly-measured probabilities ($P$) and those that are inferred ($P'$) becomes critical.

\section{Experimental Set-up \label{SEC:exp}}

Our experiment realises a quantum three-level system, or qutrit, with single photons travelling through the apparatus depicted in Fig.~\ref{FIG:Experimental_setup}.   As the general set-up is similar to previous work \cite{WCX2017}, we refer the reader to these for more details on the implementation of the various components discussed below.

The basis states of the qutrit, $|A\rangle=(1,0,0)^\text{T}$, $|B\rangle=(0,1,0)^\text{T}$, and $|C\rangle=(0,0,1)^\text{T}$, are respectively encoded by the horizontal polarization of the heralded single photons in the upper mode $\ket{HU}$, the vertical polarization of the photons in the upper mode $\ket{VU}$, and the horizontal polarization of the photons in the lower mode $\ket{HD}$.  For this experiment, the photons are prepared in the initial state $\ket{C}$.
The unitary evolution of the qutrit state is realised by a sequence of half-wave plates (HWPs) and subsequent birefringent calcite beam displacers (BDs) that realise two unitary operators $U_{21}(\theta_1,\chi_1,\phi_1)$ and $U_{32}(\theta_2,\chi_2,\phi_2)$ that are nominally identical and can be decomposed as~\cite{Reck1994,Wang2016}
\begin{align}
  &U(\theta,\chi,\phi)=\\
  &\begin{pmatrix}
  1&0&0\\
  0&\cos\theta&\sin\theta\\
  0&-\sin\theta&\cos\theta\\
  \end{pmatrix}\begin{pmatrix}
  \cos\chi&0&\sin\chi\\
  0&1&0\\
  -\sin\chi&0&\cos\chi\\
  \end{pmatrix}\begin{pmatrix}
  \cos\phi&\sin\phi&0\\
  -\sin\phi&\cos\phi&0\\
  0&0&1
  \end{pmatrix}
  .
  \nonumber
  \label{eq:U}
\end{align}

Throughout the experiment, measurement of the photon state at $t_3$ is always performed projectively.  This is accomplished by BD$_9$ that maps the basis states of qutrit into three spatial modes followed by single-photon avalanche photodiodes (APDs), in coincidence with the trigger photons.  The probability of the photons being measured in $\ket{A}$, $\ket{B}$ or $\ket{C}$ is obtained by normalizing photon counts in the certain spatial mode to total photon counts. The count rates are corrected for differences in detector efficiencies and losses before the detectors. We assume that the lost photons would have behaved the same as the registered ones (fair sampling)~\cite{Giustina2012}. Experimentally this trigger-signal photon pair is registered by a coincidence count at APD with $3$ns time window. Total coincidence counts are about $14,000$ over a collection time of $7$s.

In the forms we consider them here, the Leggett-Garg and NSIT tests require two different types of measurement of time $t_2$, i.e. between the two unitary evolution operations.
The unambiguous measurement is realized by placing blocking elements into the optical paths~\cite{WCX2017,Emary2012a}. With, for example, the channels $B$ and $C$ blocked, the joint probabilities $P(n_3, n_2 = A)$ is obtained without the measurement apparatus having interacting with the photon. In our experiment, this blocking is realized by a polarizing beam splitter (PBS) following by beam stoppers. The PBS is used to map the basis states of qutrit to three spatial modes and the beam stoppers are used to block photons in two of the three spatial modes and let the photons in the rest one pass through. By inserting the HWPs before and after the PBS, we can block any two of the channels and let the photons in the rest one pass through for the next evolution.

The ambiguous measurement is realized in a similar fashion but this time we block just one mode and let photons propagate forwards from the remaining two.  With channel $C$ blocked, for example, and with projective measurements at $t_3$, we obtain the joint probability $P(n_3,n_2=A\cup B)$, where the inference that the photon must have occupied either state $A$ or $B$ at time $t_2$ being the essential ambiguity in this scheme.

\section{LGI with unambiguous measurements \label{SEC:uLGI}}

We first consider an LGI test with unambiguous measurements.  In the case where the state preparation is elected to coincide with the first measurement \cite{Goggin2011,Robens2015,Lambert2016,WCX2017,WCX2017}, the LGI correlator reads
\beq
  K = \ew{Q_2} + \ew{Q_3 Q_2} - \ew{Q_3}
  \label{EQ:Kreducedfirst}
  .
\eeq
The expectation value $\ew{Q_3}$ is obtained using time-evolution operators $U_{21}$ and $U_{32}$ applied sequentially, followed by a projective measurement.  This yields the probabilities $P(n_3)$ and
$\ew{Q_i} = \sum_{n_i} q(n_i) P(n_i) $.  Here  the quantities  $q(A)=-q(B)=q(C)=1$ define a mapping from observed state $n$ to dichotomic variable $Q$ \cite{Budroni2014}.  The remaining correlation functions are obtained as $\ew{Q_3 Q_2} = \sum_{n_3,n_2} q(n_3) q(n_2) P(n_3,n_2)$
and
$\ew{Q_2} = \sum_{n_3,n_2} q(n_2) P(n_3,n_2)$ with the joint probabilities $P(n_3,n_2)$ being obtained from experimental runs in which evolution operators $U_{21}$ and $U_{32}$ have projective measurements situated both between and after them.

Under the standard LG assumptions, this correlator obeys
$
  K \le 1
$.
However, we will consider the form of the LGI given in Ref.~\cite{Clive2017}, which avoids the non-invasive-measureability assumption.  In this case, we obtain the ``modified LGI'', which reads
\beq
  K\leq 1+\Delta,\quad \Delta\equiv\sum_{n_3}|\delta(n_3)|
  \label{EQ:modLGI}
  .
\eeq
Here
\beq
  \delta(n_3) = P(n_3) - \sum_{n_2} P(n_3,n_2)
  \label{EQ:firstdeltadefn}
  ,
\eeq
describes the amount of signalling from time $t_2$ to $t_3$.  Under assumption of non-invasive measureability, this would be zero such thats we have
\beq
  \delta(n_3) =0;~\forall n_3.
  \label{EQ:NSIT}
\eeq
These are the NSIT equalities \cite{Kofler2013} and if they are satisfied, the modified LGI reduces to its original form $K\le 1$.
In the approach we pursue here, however, we take the quantities $\delta(n_3)$ to be obtained from experiment, and consider the modified LGI, \eq{EQ:modLGI}, in this light.
\begin{figure}[tb]
   \includegraphics[width=\columnwidth,clip=true]{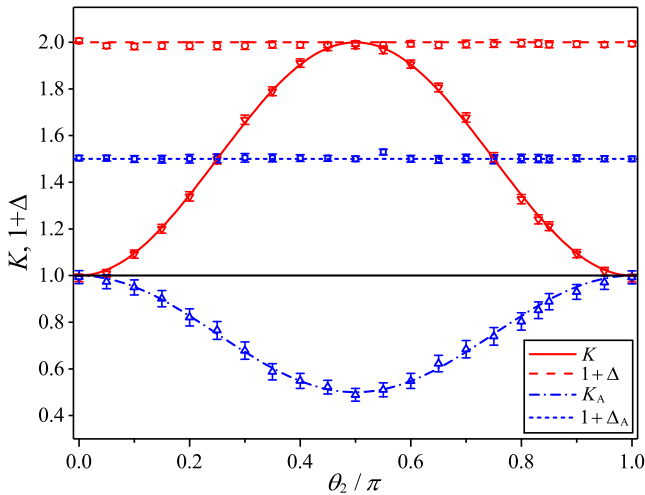}
   \caption{
      Experimentally-determined values of the LG correlator $K$ and the corresponding right-hand side $1+\Delta$ of the modified LGI, \eq{EQ:modLGI} with evolution parameters $\theta_1=0,\chi_1=\chi_2-\pi/2=\pi/4,\phi_1=\phi_2=0$ and a range of  $\theta_2$ values.  These parameters are chosen to maxmize the value of the unambiguously-measured $K$.  However,  although we have $K\ge1$ for all values of $\theta_2$, we have $K\le1+\Delta$ throughout the tested range.  Thus when the observed signalling taken into account, the modified LGI, \eq{EQ:modLGI}, is never violated.
      For completeness, we also plot the ambiguously-measured $K_\text{A}$  and $1+\Delta_\text{A}$ for the same parameters. Here too we observe $K_\text{A}\le1\le1+\Delta_\text{A}$ and no violations are recorded.
      Theoretical predictions are represented by curves and lines, and the experimental results by symbols.
      Error bars include both the statistical uncertainty and the error due to the inaccuracy of the wave plate alignment.
      \label{FIG:data}
   }
\end{figure}

Figure~\ref{FIG:data} shows a comparison of the measured values of $K$ and $1+\Delta$ from our experiment with unambiguous measurements.  We have selected a particular choice of evolution operators such that, for these parameters, we find analytically that
$K=(3-\cos2\theta_2)/2$, with $\theta_2$ an adjustable evolution parameter.  As Fig.~\ref{FIG:data} shows, this result is very closely matched by experiment.
Error bars in this figure include both the statistical uncertainty and the error due to the inaccuracy of the wave plates~\cite{Xu2014}.  The statistical errors based on the assumption of Poissonian statistics are relatively small.  However, about $20$ wave plates are used and each of them has an angle error of approximately $0.1^\circ$.  These errors accumulate in a cascaded setup and we have simulated numerically their total effect with a Monte Carlo method.  These inaccuracies are sufficient to explain deviations from theoretical predictions.

The maximum value of $K$ that we observe is $1.988\pm 0.016$ at $\theta_2 = \pi/2$, which agrees well with the theoretical prediction of $2$.
This value represents an enhanced violation of the LGI, above the bound set for genuinely-dichotomic measurements, as described in \cite{Budroni2014} and observed experimentally in \cite{WCX2017}. It is clearly far in excess of the usual LGI macrorealistic bound of $K\le1$.
For these parameters, we obtain the righthand side of the LGI as
 $1+\Delta = 2 $ analytically, which is constant as a function of $\theta_2$.  This behaviour is recovered by experiment and for $\theta_2 = \pi/2$ we obtain $1+\Delta =1.995 \pm 0.011$.
Thus, whilst the observed value of $K$ is clearly in excess of the standard bound, when the observed degree of signalling taken into account, we find that the modified LGI, \eq{EQ:modLGI}, still holds. This is line with the theoretical results of Refs.~\cite{George2013,Clive2017} which forbid violations of \eq{EQ:modLGI} with projective measurements.

\begin{figure}[tb]
   \includegraphics[width=\columnwidth,clip=true]{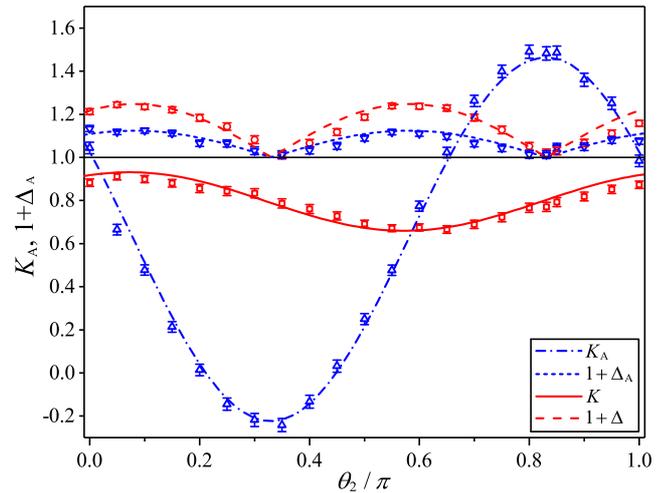}
   \caption{
     Experimentally-determined values of LG correlator and upper bound for a second set of parameters: $\theta_1=0.831\pi,\chi_1=\chi_2=0.688\pi,\phi_1=\phi_2=0.423\pi$)
     and a range of values of $\theta_2$.
     In this case, focus is on the ambiguously measured correlator $K_\text{A}$ and its bound $1+\Delta_\mathrm{A}$.
     Results are also shown for the unambiguously-measured $K$ and $1+\Delta$.
     For the $\theta_2 = 0.831\pi$, we observe a value of $ K_\text{A} = 1.483\pm0.031$ whilst both $1+\Delta_\mathrm{A}$ and $1+\Delta$ close to 1 to within. At this point, then, we observe LGI violations in the absence of signalling for both measurement types.
     Other details as in \fig{FIG:data}
     \label{FIG:data2}
   }
\end{figure}

\section{LGI with ambiguous measurements \label{SEC:aLGI}}

Following Ref.~\cite{Clive2017}, an LGI constructed with the ambiguous measurements has exactly the same form as before
\begin{equation}
   K_\text{A}\leq1+\Delta_\text{A},
   \quad
   \Delta_\text{A}=\sum_{n_3}|\delta_\text{A}(n_3)|
   \label{EQ:modaLGI}
   ,
\end{equation}
where the subscript A denotes quantities obtained from ambiguous measurements.  These quantities have forms identical to those considered previously but with probabilities $P(n_3,n_2)$ replaced with those inferred from the ambiguous measurements.  In particular, we obtain the joint probabilities $P'(n_3,A)$ in the same way as \eq{EQ:inferPA} and write
\beq
  P'(n_3,A) &=&
  \textstyle{\frac{1}{2}}P(n_3,A\cup B)
  + \textstyle{\frac{1}{2}}P(n_3, A\cup C)
  \nonumber\\
  &&
  -\textstyle{\frac{1}{2}}P(n_3,B\cup C)
  \label{EQ:inferPnA}
  ,
\eeq
and similarly for the other two probabilities.  The ambiguously-measured no-signaling quantities are then
$
  \delta_\text{A}(n_3) \equiv  P(n_3)-\sum_{n_2}P'(n_3,n_2)\nonumber
\label{eq:ESIT}
$, and the correlation functions in $K_\text{A}$ are the same as before with the replacement $P \to P'$.
The ambiguously-measured probabilities $P(n_3,\alpha)$ are obtained experimentally in exactly the same way as before, but with ambiguous measurements replacing the unambiguous one at $t_2$.  Theoretically, they are obtained with a POVM as outlined in Ref.~\cite{Clive2017}. Note that the algebraic bound of $K=3$ is never violated, irrespective of measurement type \cite{Dressel2011,Budroni2014}.

Results for $K_\text{A}$ and $1+\Delta_\text{A}$ for the parameter set in Sec.~\ref{SEC:uLGI} are shown in \fig{FIG:data}.  In this case $K_\text{A}\le1\le1+\Delta_\text{A}$ and no violations of the ambiguously-measured LGI are observed.

Figure~\ref{FIG:data2}, however, shows these quantities for a different set of evolution parameters, namely $\theta_1=0.831\pi,\chi_1=\chi_2=0.688\pi,\phi_1=\phi_2=0.423\pi$ and $0\leq\theta_2\leq \pi$. For a significant range of $\theta_2$ values, we obtain $K_\mathrm{A}>1$. Moreover, for $0.677\pi \le \theta_2 \le0.983\pi$, we find that $K_\text{A}\geq1+\Delta_\text{A}$, and thus we find violations of the modified ambiguously-measured LGI.  The maximum violation is found at $\theta_2=0.831\pi$ with values $ K_\text{A} = 1.483\pm0.031$, in close agreement with the theoretical prediction $1.464$.
Most importantly, at this value of $\theta_2$ the signalling quantities are $\Delta=0.019\pm0.020$ and $\Delta_\text{A}=0.013\pm0.018$ both of which are, to within experimental uncertainty, essentially zero in accordance with theory which gives $\Delta=\Delta_\text{A}=0$ exactly at this point.

\begin{figure}[tb]
   \includegraphics[width=\columnwidth,clip=true]{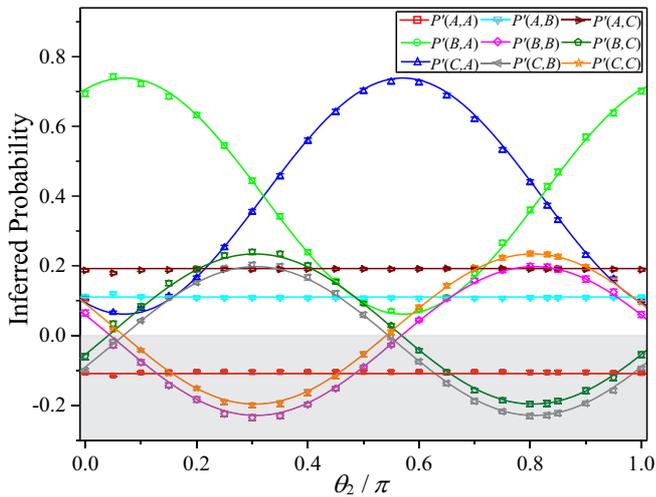}
   \caption{
     Experimentally-determined values of the inferred joint probabilities $P'(n_3,n_2)$ (for $n_2,n_3 = A,B,C$) as a function of the parameter $\theta_2$ (the other parameters fixed as in \fig{FIG:data2}). Theoretical predictions are represented by curves and lines, and the experimental results by symbols. That $P'$ takes on negative values is indicative of the quantum-mechanical quasi-probabilistic nature of these quantities.
     \label{FIG:data3}
   }
\end{figure}

Whilst our experiment therefore satisfies no-signalling for both sets of measurements, and therefore also shows equality of signalling between them, we can understand the origins of the LGI violations in our scheme by looking at the individual inferred probabilities $P'(n_3,n_2)$.  The complete set of these are plotted in Fig.~\ref{FIG:data3} for the parameters of Fig.~\ref{FIG:data2}.  Crucially, for all values of $\theta_2$ at least three of inferred probabilities are \textit{negative}. For example $P'(A,A)=-0.109$ for all $\theta_2$.

These results make it clear that, in quantum-mechanical terms, these inferred  probabilities are quasi-probabilities.  The role of negative quasi-probabilities in violations of LGIs has been discussed a number of times~\cite{Calarco1999,Suzuki2012,Clive2015,Halliwell2016,Halliwell2017}.
In addition, anomalous weak values have been directly connected to LGI violations~\cite{Dressel2011,PRL08}, and such values imply the existence of negative quasi-probabilities~\cite{PRA15}.  The link between negative quasi-probabilities and contextuality reported in Refs.~\cite{PRL14,PRA17} also implies a connection between the LGI violations and contextuality.

\section{Discussion\label{SEC:disc}}

We have described here the experimental violation of the LGI using a realisation of a three-level system with single photons.   We have shown that it is possible to obtain violations of the modified LGI, \eq{EQ:modaLGI}, that takes into account the observed degree of signalling. Violations of this inequality were observed for a range of our evolution parameter $\theta_2$.
Moreover, at the particular point $\theta_2 = 0.831\pi$, both signalling quantities $\Delta$ and $\Delta_\text{A}$ were found to be zero.
At this point, then, NSIT is obeyed by both the ambiguous and unambiguous measurements.  This is particularly important because, according to Ref.~\cite{Clive2017}, the derivation of \eq{EQ:inferPnA} and hence \eq{EQ:modaLGI} relies on the assumption that both unambiguous and ambiguous measurements  are ``equally invasive'' and therefore must exhibit the same degree of signalling, i.e. $\Delta=\Delta_\text{A}$ \footnote{This so-called \textit{equal signalling in time} condition can be seen by summing  \eq{EQ:inferPnA} over $n_2$}.
Only at the point $\theta_2 = 0.831\pi$, are the dynamics of our three-level system such that we have $\Delta=\Delta_\text{A}$.  At this point then the use of the ambiguous measurements to construct the LGI for ``macrorealistic'' state $n_i$ is justified.

Due to its use of photons, this is a proof-of-principle experiment and can not be viewed as a test of \textit{macroscopic} realism, as originally envisaged by Leggett and Garg but rather of \textit{microscopic} realism \cite{Bohm1985,Leggett2008} as has been famously tested in Bell-type experiments \cite{Genovese2005}. Nevertheless the general principle used for constructing ambiguous LGI tests without signalling could potentially be scaled up to larger, massive objects, perhaps most directly in molecular interference experiments \cite{Emary2014b}.

Despite the enhanced no-signalling features of our experiment, and in common with all known Leggett-Garg-type tests, possible loopholes exist for a macrorealist determined to hold their position.  The finding that some of the the inferred probabilities, $P'(n_3,n_2)$ are negative would presumably lead the macrorealist to reject the possibility that it possible to learn anything about the unambiguous state of the system from ambiguous set-up. This position, however, would require a significant degree of contrivance given that both measurements are known to be  individually non-signalling.

\begin{acknowledgments}
We are grateful to J.~J.~Halliwell for suggesting analysis of the individual quasiprobabilities.
We acknowledge support by NSFC (Nos.~11474049 and 11674056), NSFJS (No. BK20160024), the Open Fund from State Key Laboratory of Precision Spectroscopy, East China Normal University and the Scientific Research Foundation of the Graduate School of Southeast University.
\end{acknowledgments}

\bibliography{ambiguous}
\end{document}